\documentclass{PoS}
\usepackage{amsmath}
\newcommand{\I}{\ensuremath{\mathrm{i}}}
\newcommand{\E}{\ensuremath{\mathrm{e}}}
\newcommand{\tr}{\ensuremath{\mathrm{tr}}}
\newcommand{\arxiv}[1]{arXiv:\,\href{http://arxiv.org/abs/#1}{{\tt #1}}}
\newcommand{\aetap}{\text{a--}\eta'}
\newcommand{\api}{\text{a--}\pi}
\newcommand{\afn}{\text{a--}f_0}
\newcommand{\glg}{\tilde{g}g}
\title{$\mathcal{N}$=1 supersymmetric Yang-Mills theory on the lattice}

\ShortTitle{$\mathcal{N}$=1 supersymmetric Yang-Mills theory on the lattice}

\author{Georg Bergner\\
University of Frankfurt, Institute for Theoretical Physics,\\
Max-von-Laue-Str.~1, D-60438 Frankfurt am Main, Germany}
\author{Istvan Montvay\\
Deutsches Elektronen-Synchrotron DESY,\\
Notkestr. 85, D-22603 Hamburg, Germany}
\author{\speaker{Gernot M\"unster}, Dirk Sandbrink, Umut D.\ \"Ozugurel\\
University of M\"unster, Institute for Theoretical Physics,\\
Wilhelm-Klemm-Str.~9, D-48149 M\"unster, Germany\\
E-mail: \email{munsteg@uni-muenster.de}}

\abstract{Numerical simulations of supersymmetric theories on the lattice
are intricate and challenging with respect to their theoretical foundations
and algorithmic realisation. Nevertheless, the simulations of a
four-dimensional supersymmetric gauge theory have made considerable
improvements over the recent years. In this contribution we summarise the
results of our collaboration concerning the mass spectrum of this theory.
The investigation of systematic errors allows now a more precise estimate
concerning the expected formation of supersymmetric multiplets of the
lightest particles. These multiplets contain flavour singlet mesons,
glueballs, and an additional fermionic state.}

\FullConference{31st International Symposium on Lattice Field Theory
LATTICE 2013\\
July 29 - August 3, 2013\\
Mainz, Germany}
\begin{document}

\section{Introduction}

Supersymmetry (SUSY) is a central concept in models for physics beyond the
Standard Model of elementary particle physics. Supersymmetric theories also
provide us with intriguing theoretical structures. Supersymmetry is an
extension of the Poincar{\'e} symmetry of space-time corresponding to an
extension of the Poincar{\'e} algebra by one or several supersymmetry
charges to a {\em super-Poincar{\'e} algebra} that relates bosons to
fermions. The supercharges change the spin by $\frac{1}{2}$, hence the
supersymmetry multiplets contain particles with different spins, in
particular bosons and fermions, at the same time. Since at present energies
no such supermultiplets with degenerate masses are observed, supersymmetry
-- if it is realised in Nature -- has to be a broken symmetry.

Part of supersymmetric extensions of the Standard Model is the
$\mathcal{N}=1$ supersymmetric Yang-Mills theory (SYM). It is the
supersymmetric extension of Yang-Mills theory. In addition to the gluons it
contains their superpartners, the gluinos. Gluinos are Majorana fermions in
the adjoint representation of the gauge group. The Lagrangian of SYM in the
continuum is
\begin{equation}
\mathcal{L}=\tr\left[-\frac{1}{4} F_{\mu\nu}F^{\mu\nu}
+\frac{\I}{2}\bar{\lambda}\gamma^\mu
D_\mu\lambda{-\frac{m_g}{2}\bar{\lambda}\lambda} \right ], 
\end{equation}
where $F_{\mu\nu}$ is the non-Abelian field strength formed out of the gauge
fields $A_{\mu}(x)$, $\lambda(x)$ is the gluino field, and $D_\mu$ denotes
the gauge covariant derivative in the adjoint representation. The gluino
mass term breaks the supersymmetry of the theory softly. For vanishing
gluino mass it is expected that supersymmetry is not broken in SYM in the
continuum \cite{Witten:1982df}.

$\mathcal{N}=1$ SYM is similar to QCD in various
respects \cite{Amati:1988ft}, the difference being the Majorana nature of the
fermions and their colour representation. As in QCD the gauge coupling in
SYM is asymptotically free at high energies and becomes strong in the
infrared limit. Due to confinement the spectrum of particles is expected to
consist of colourless bound states. If supersymmetry is unbroken the
particles should belong to mass degenerate SUSY multiplets.

Many predictions concerning the properties of SYM are based on perturbation
theory or semiclassical methods. The low-energy properties, in particular
the structure of the particle multiplets are, however, of a non-perturbative
nature. It would therefore be desirable to study whether the theory exists
as a continuum limit of a non-perturbatively defined model regularised on a
space-time lattice, and whether the predictions about its physical
properties can be confirmed. In particular, to verify the formation of
supermultiplets is a central task for non-perturbative studies in the
lattice regularisation. Other non-perturbative properties of interest
include the spontaneous breaking of chiral symmetry, $Z_{2 N_{c}} \to
Z_{2}$, that manifests itself in the non-vanishing vacuum expectation value
$\langle\lambda\lambda\rangle \neq 0$, the confinement of static quarks,
indicated by a linear rise in the static quark potential, which is an
evidence for the confining nature of the theory, and the validity of the
supersymmetric Ward identities.

Results on these questions have been obtained by our collaboration in
previous studies in the framework of a lattice-regularised version of SYM.
Presently we study SYM with gauge group SU(2). Our non-perturbative studies
are concentrating on the properties of the light particle spectrum. In
particular, we determine the masses of the lightest composite particles by
performing numerical simulations. For our recent publications see
Refs.~\cite{Demmouche:2010sf,Bergner:2011wf,Bergner:2012rv,Bergner:2012eg,%
Bergner:2013nwa}.

\section{SYM on the lattice}

The lattice discretisation in general breaks supersymmetry
\cite{Bergner:2009vg}. Models that respect part of an extended superalgebra
are discussed in Ref.~\cite{Catterall:2009it}. For $\mathcal{N}=1$ SYM
supersymmetry is broken in any known lattice formulation. Our investigations
are based on the Curci-Veneziano lattice action \cite{Curci:1986sm}, which
is built in analogy to the Wilson action of QCD \cite{Wilson:1974sk} for the
gauge field (``gluon'') and Wilson fermion action for the gluino. It is
given by
\begin{equation}
\label{action}
S = S_g + S_f .
\end{equation}   
Here $S_g$ is the gauge field action
\begin{equation}
S_g  =   \beta \sum_{pl}
\left( 1 - \frac{1}{N_c} {\rm Re}\,\tr\, U_{pl} \right),
\end{equation}
with the gauge coupling $\beta \equiv 2N_c/g^2$ for an SU($N_c$) gauge
field. $U_{pl}$ is the product of the gauge link fields along a plaquette.
In our simulations we actually use the tree-level Symanzik improved gauge
action, which contains, besides the plaquettes, also rectangular Wilson
loops of perimeter six. The fermionic part of the action (\ref{action}) is
\begin{equation}
\begin{array}{rcl}
S_f &\equiv& \frac{1}{2} \overline{\lambda} Q \lambda \\[2mm]
&\equiv& \frac{1}{2} \sum_x \left\{ \overline{\lambda}_x^a\lambda_x^a
-K \sum_{\mu=1}^4 \left[
\overline{\lambda}_{x+\hat{\mu}}^a V_{ab,x\mu}(1+\gamma_\mu)\lambda_x^b
+\overline{\lambda}_x^a V_{ab,x\mu}^T (1-\gamma_\mu)
\lambda_{x+\hat{\mu}}^b \right] \right\}.
\end{array}
\end{equation}
Here $K$ is the hopping parameter which determines the gluino mass,
$\gamma_\mu$ denotes a Dirac matrix and $V_{x\mu}$ is the gauge field
variable in the adjoint representation of the gauge group, which is obtained
from the gauge field links in the fundamental representation $U_{x\mu}$ by
\begin{equation}
V_{x\mu}^{ab} \equiv  
2 \,\tr \left( U^\dagger_{x\mu} T^a  U_{x\mu} T^b \right)
\end{equation}
($T^a$ are the generators of SU($N_c$)). The gluino field $\lambda_x$
satisfies the Majorana condition
\begin{equation}
\overline{\lambda}_x = \lambda_x^T \, C
\end{equation}
with the charge conjugation Dirac matrix $C$. In the simulations we apply
one or three levels of stout smearing to the link fields in the Wilson-Dirac
operator.

Performing the path integral over the fermion field $\lambda$ results in a
{\em Pfaffian}:
\begin{equation}
\int [d\lambda] \E^{ -\frac{1}{2}\overline{\lambda} Q \lambda }
= \int [d\lambda] \E^{ -\frac{1}{2}\lambda M \lambda } = {\rm Pf}(M),
\end{equation}
where $M$ is the antisymmetric matrix defined as
\begin{equation}
M \equiv CQ = -M^T .
\end{equation}
The square of the Pfaffian ${\rm Pf}(M)$ is equal to the determinant of the
fermion matrix $Q$:
\begin{equation}
\det(Q) = \det(M) = \left[ {\rm Pf}(M) \right]^2 \,.
\end{equation}
The lattice discretisation breaks both supersymmetry and chiral symmetry,
but they are expected to be restored in the continuum limit if the hopping
parameter is tuned to a critical value $K_c$ corresponding to a vanishing
gluino mass. The breaking of chiral symmetry and the related need to tune
the hopping parameter could be avoided by using domain-wall
\cite{Giedt:2008xm,Endres:2009yp} or overlap \cite{Kim:2011fw} fermions, but
the SUSY breaking remains and the required numerical effort for simulations
would substantially increase.

In the continuum the Pfaffian introduced above is positive, but on the
lattice this is not necessarily the case. Therefore in the numerical
simulations its sign has to be taken into account separately. Taking the
non-negative square root of the determinant, the effective gauge field
action is \cite{Curci:1986sm}:
\begin{equation}
S_{CV} =
\beta\sum_{pl} \left( 1 - \frac{1}{N_c} {\rm Re}\,\tr\,U_{pl} \right)
- \frac{1}{2}\log\det Q[U] \,.
\end{equation}
The factor $\frac{1}{2}$ in front of $\log\det Q$ corresponds to a flavour
number $N_f=\frac{1}{2}$ of Dirac fermions. The gauge configuration for this
fractional flavour number can be created, for instance, by the {\em two-step
polynomial Hybrid Monte Carlo (TSPHMC)} algorithm \cite{Montvay:2005tj},
which is our choice for Monte Carlo updating. The omitted sign of the
Pfaffian can be taken into account by reweighting:
\begin{equation}
\langle A \rangle = \frac{\langle A\; {\rm sign Pf}(M)\rangle_{CV}}
{\langle {\rm sign Pf}(M)\rangle_{CV}} \,,
\end{equation}
where $\langle \ldots \rangle_{CV}$ denotes expectation values with respect
to the effective gauge action $S_{CV}$. This reweighting could in principle
lead to a {\em sign problem} if a strong cancellation occurs among
contributions with opposite sign. In previous work
\cite{Kirchner:1998mp,Campos:1999du} we have shown by monitoring the sign of
the Pfaffian that for positive gluino masses the positive contributions
dominate and there is practically no sign problem.

\section{Light particle spectrum}

The main interest of our present investigations of SYM is in the low-lying
spectrum of particles. Colour neutral bound states can be created from the
vacuum by gauge invariant operators which are built from the gluon and
gluino field operators. These include the adjoint mesons $\afn$ and
$\aetap$. They are composed of two gluinos with spin-parity $0^+$ and $0^-$,
respectively. The corresponding bilinear operators are $\bar\lambda \lambda$
and $\bar\lambda \gamma_5 \lambda$. The meson propagator has connected and
disconnected contributions:
\begin{equation}
\begin{split}
C_{\Gamma}(t)
&= \frac{1}{V_s} \sum_{\vec x,\vec y} \left\langle
\underbrace{\tr_{sc}[\Gamma Q_{xx}^{-1}]
\tr_{sc}[\Gamma Q_{yy}^{-1}]}_{\mbox{disconnected}}
- 2 \underbrace{\tr_{sc}[\Gamma Q_{xy}^{-1}
\Gamma Q_{yx}^{-1}]}_{\mbox{connected}} \right\rangle\\
&- \frac{1}{V_s} \left\langle \frac{1}{T} \sum_t \sum_{\vec x}
\tr_{sc} [\Gamma {Q_{xx}^{-1}}]\right\rangle^2\!\!,
\end{split}
\end{equation}
where $\tr_{sc}$ denotes a trace over spin and colour indices, and $\Gamma =
1$ or $\gamma_5$, respectively. The numerical evaluation of the disconnected
propagators is rather demanding. In order to reduce the large variance, the
disconnected part has been calculated using the stochastic estimator method
\cite{Bali:2009hu} combined with a truncated eigenmode approximation. As it
is the case in QCD, the disconnected diagrams are intrinsically noisier than
the connected ones and dominate the level of noise in the total correlator.

Other particles which are expected to exist in this model are the glueballs,
which are created by purely gluonic operators. In particular, one expects
low-lying $0^+$ and $0^-$ glueballs. In addition to scalar and pseudoscalar
particles, a chiral supermultiplet would contain a Majorana fermion
particle. Such particles are provided by the {\em gluino-glueballs}. They
are represented by the operator
\begin{equation}
\tilde{O}_{g\tilde{g}} =
\sum_{\mu\nu} \sigma_{\mu\nu} \tr \left[ F^{\mu\nu}  \lambda\right],
\end{equation}
where $\sigma_{\mu\nu}=\frac{1}{2} \left[ \gamma_\mu,\gamma_\nu \right]$ and
$F^{\mu\nu}$ is the field strength tensor. A lattice version of this, which
can be used in numerical simulations, is
\begin{equation}
O^\alpha_{g\tilde{g}}= \sum_{i<j, \beta} \sigma^{\alpha\beta}_{ij}\,
\tr \left[ P_{ij} \lambda^\beta \right] \, ,
\end{equation}
where the indices $i$ and $j$ stand for the spatial directions. A choice for
$F_{ij}$ with the proper parity and time reversal transformation properties
is the antihermitian part of the clover plaquette $U^{(c)}$
\begin{equation}
P_{ij}=\frac{1}{8 \I g_0} (U^{(c)}_{\mu\nu}-(U^{(c)}_{\mu\nu})^\dag)\, .   
\end{equation}   
For its definition and more details see Ref.~\cite{Bergner:2012rv}.

On the basis of effective Lagrangeans, in
Refs.~\cite{Veneziano:1982ah,Farrar:1997fn} it has been predicted that the
low-lying particles form two chiral supermultiplets, each consisting of a
scalar, a pseudoscalar, and a fermionic spin 1/2 particle. One of them
contains an $\aetap$, an $\afn$, and a gluino-glueball, the other one a
$0^-$ glueball, a $0^+$ glueball, and a gluino-glueball.

In our numerical investigations we have calculated the masses of the
gluino-glueball, $\aetap$, $\afn$, and $0^+$ glueball. In addition the mass
of the adjoint pion ($\api$) is obtained. The correlator of this particle is
given by the connected contribution of the $\aetap$ correlator. The $\api$
is not a physical particle in SYM. However, on the basis of arguments
involving the OZI-approximation of SYM \cite{Veneziano:1982ah}, the adjoint
pion mass is expected to vanish for a massless gluino and the behaviour
$m^2_{\api} \propto m_g$ can be assumed for light gluinos. The corresponding
value of $K_c$ is most easily obtained from the dependence of the
$\api$-mass on $K$.

The gluino-glue correlator has been obtained using different smearing
techniques. The link fields are smeared using APE smearing, the fermionic
fields using Jacobi smearing. In order to decrease lattice artefacts and
statistical fluctuations in the Wilson-Dirac fermion matrix $Q$ of the
lattice action, the gauge link variables $U_{x\mu}$ have been replaced by
{\em stout smeared} links \cite{Morningstar:2003gk}.

In Ref.~\cite{Bergner:2012rv} a detailed investigation of finite size
effects was done, showing that the lattice volumes used in our simulations
are sufficiently large, such that the finite volume effects are smaller than
the statistical errors.

The simulations are performed at non-zero gluino masses where supersymmetry
is softly broken. The obtained masses are then extrapolated to vanishing
gluino mass at $K=K_c$. The following figures show the masses of the
particles discussed above as a function of the squared mass of the adjoint
pion, which is a convenient substitute for the gluino mass. The inverse bare
gauge coupling in the current simulations was $\beta=1.75$. The scale is set
by the Sommer parameter $r_0$, obtained from the static quark potential. We
use QCD units by setting the Sommer parameter to $r_0=0.5 \,\mathrm{fm}$.
The lattice spacing is approximately $0.055$ fm for 1 level, and
approximately $0.058$ fm for 3 levels stout smearing. Also shown in the
figures are the extrapolations to the limit of vanishing gluino mass. All
figures include the gluino-glue mass for comparison.

\begin{figure}[t]
\begin{center}
\includegraphics[width=62mm]{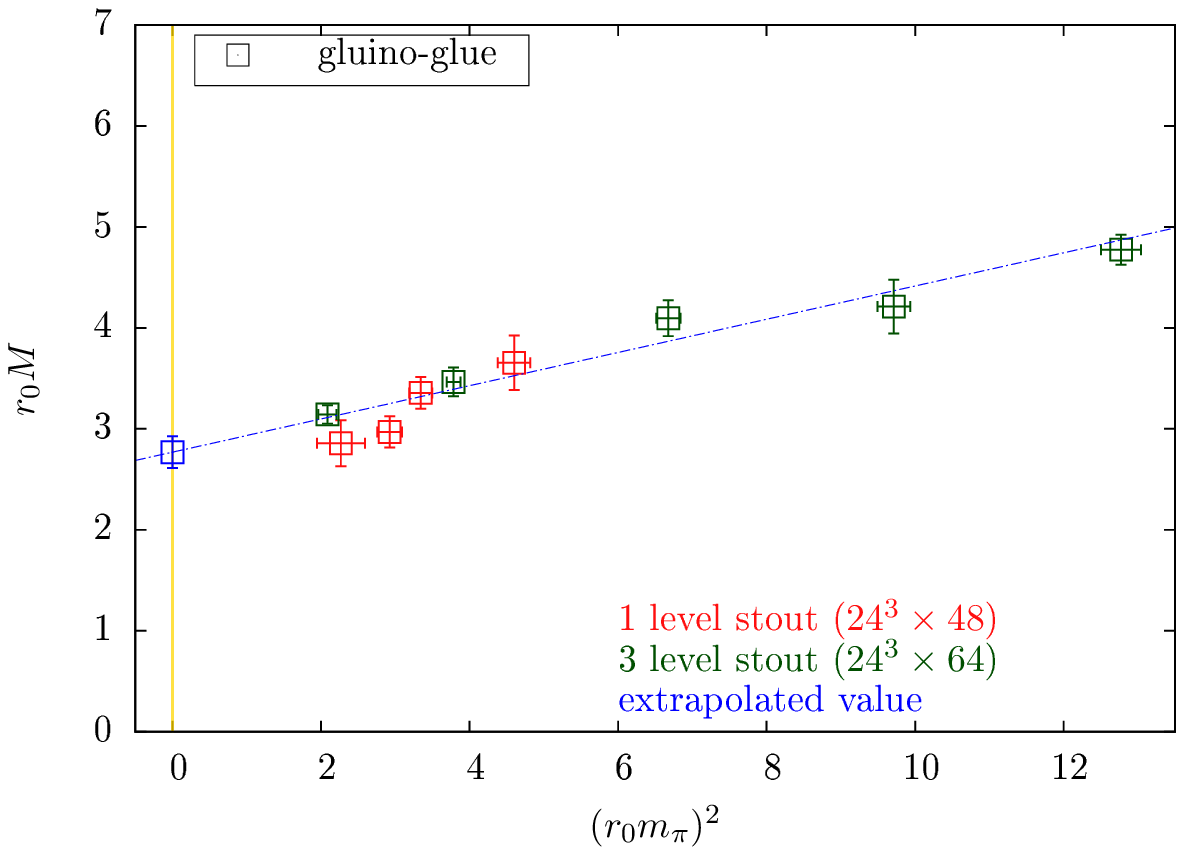}
\includegraphics[width=62mm]{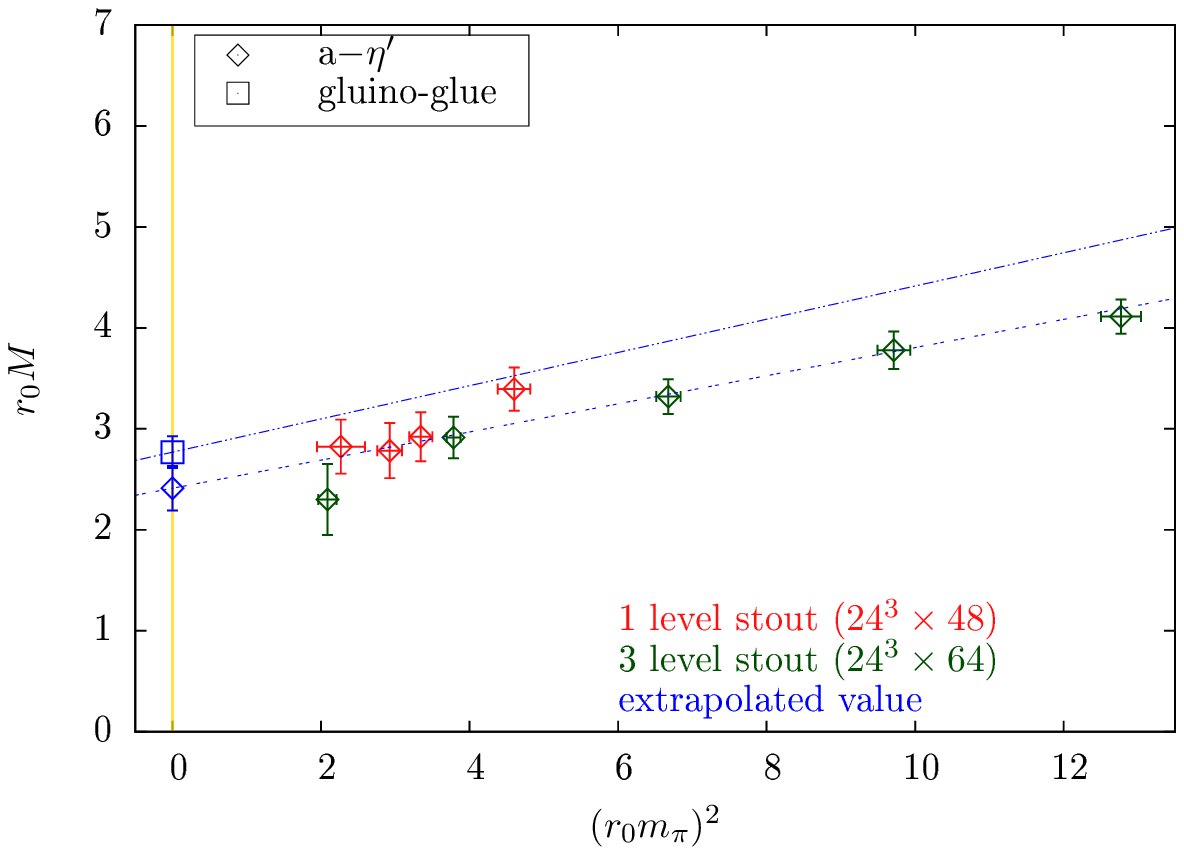}
\end{center}
\vspace*{3mm}
\begin{center}  
\includegraphics[width=62mm]{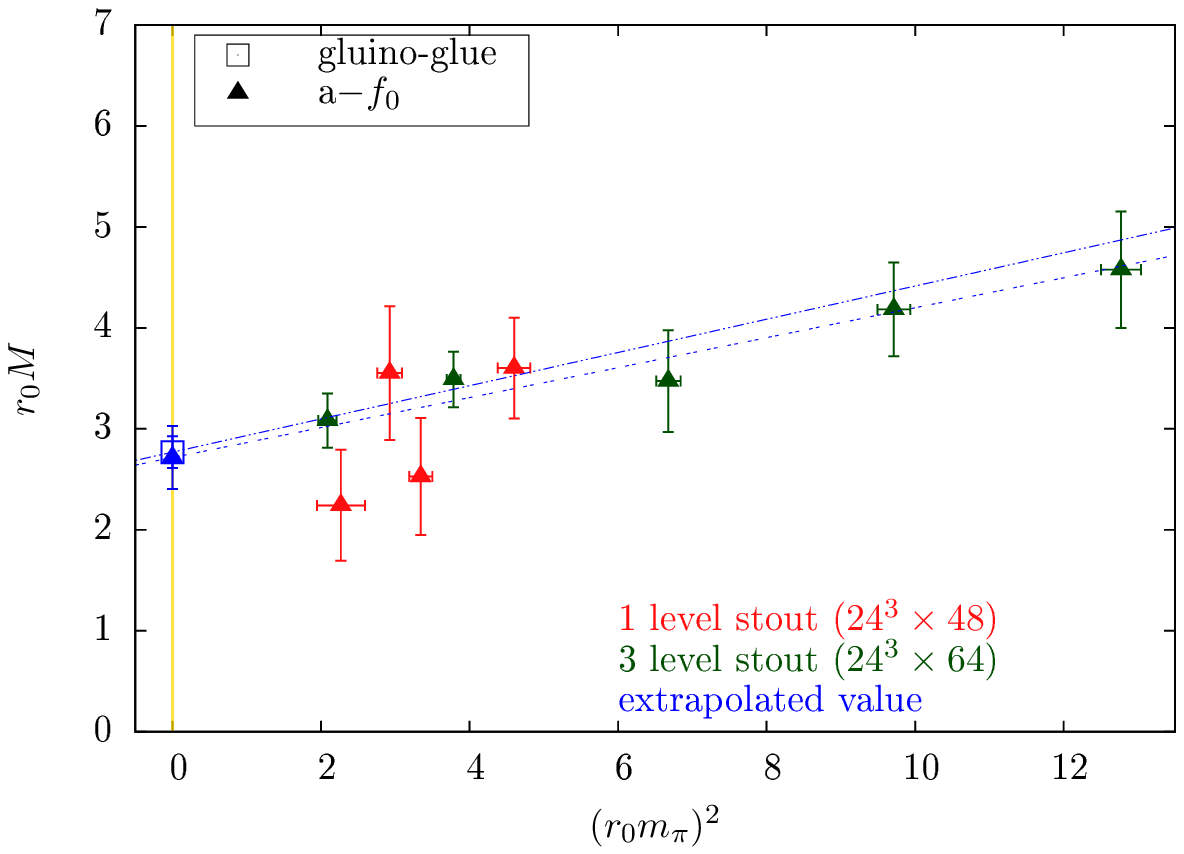}
\includegraphics[width=62mm]{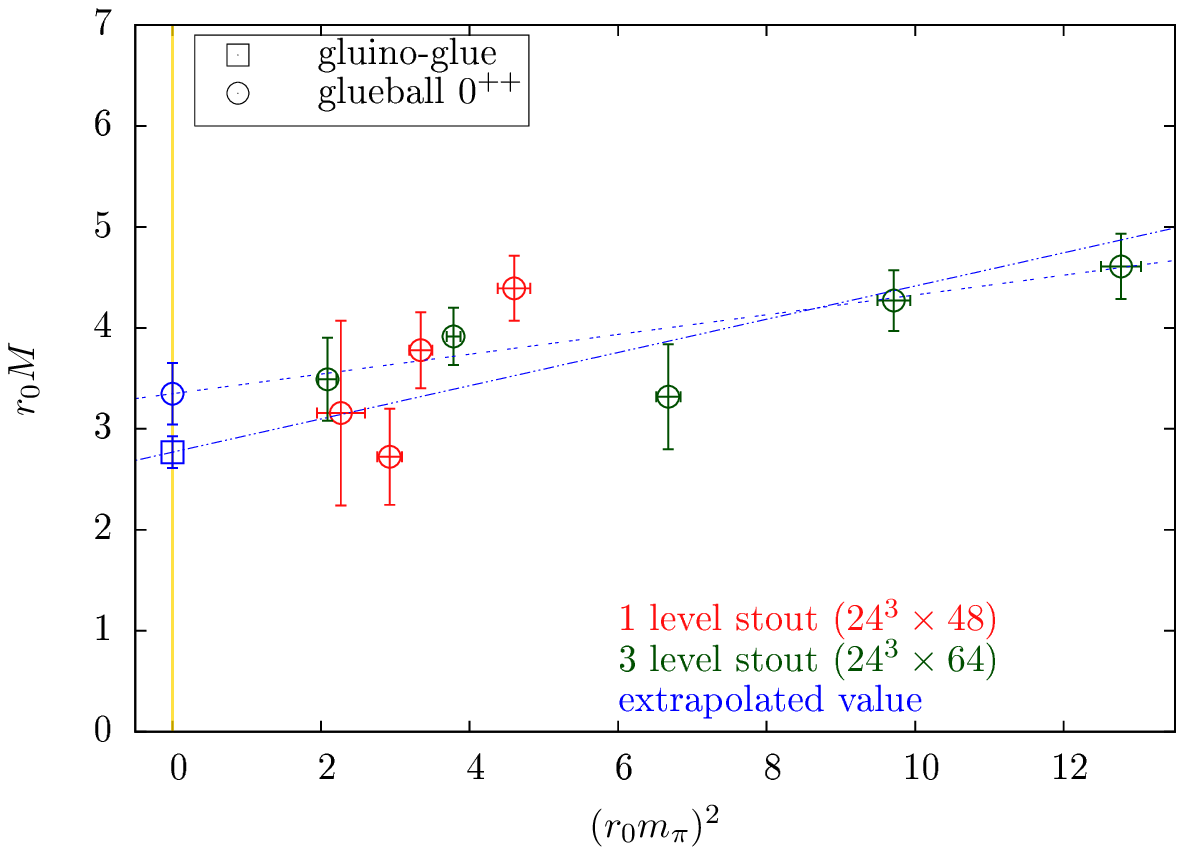}
\end{center}
\caption{\label{muenster-masses} 
Light particle masses in SYM as a function of the squared mass of
the adjoint pion. $r_0$ is the Sommer scale parameter.
}
\end{figure}

\begin{table}[htb]
\begin{center}
\begin{tabular}{l|crcr}
\hspace{1mm} $\beta$ & $\aetap$ & $\afn$\hspace{4mm} & $\glg$ 
& glueball $0^{++}$\hspace*{-3mm}\\
\hline
1.75 & 950(87) & 1070(123) & 1091(62) & 1319(120) 
\end{tabular}
\caption{Bound state masses in units of MeV, extrapolated
to vanishing gluino mass.}
\label{tab:final_res}
\end{center}
\end{table} 

The values of the extrapolated masses in QCD units are given in Table
\ref{tab:final_res}. In contrast to previous results, which were afflicted
by larger systematic errors, our recent results are consistent with the
emergence of a mass-degenerate chiral supermultiplet.

This is an important indication that this supersymmetric theory can be
simulated on the lattice and nontrivial non-perturbative results are
consistent with the theoretical prediction of an absent spontaneous
supersymmetry breaking \cite{Witten:1982df}.

\section*{Acknowledgements}

This project is supported by the German Science Foundation (DFG) under
contract Mu 757/16.
The authors gratefully acknowledge the computing time granted by the John
von Neumann Institute for Computing (NIC) and provided on the supercomputers
JUQUEEN and
JUROPA at J\"ulich Supercomputing Centre (JSC).
Further computing time has been   
provided by the compute cluster PALMA of the University of M\"unster.


\end{document}